\documentclass[useAMS]{mn2e}
\usepackage{graphicx}

\begin{document}

\title[The two hybrid B-type pulsators $\nu$~Eri and 12~Lac]
{The two hybrid B-type pulsators: $\nu$~Eridani and
12~Lacertae}
\author[W. A. Dziembowski and A. A. Pamyatnykh]
  {W. A. Dziembowski$^{1,2}$\thanks{E-mail: wd@astrouw.edu.pl}
   and A. A. Pamyatnykh$^{1,3}$
\and \\
$^{1}$ Copernicus Astronomical Center, Bartycka 18, 00-716 Warsaw, Poland\\
$^{2}$ Warsaw University Observatory, Al. Ujazdowskie 4, 00-478 Warsaw,
Poland\\
$^{3}$ Institute of Astronomy, Russian Academy of Sciences,
Pyatnitskaya Str. 48, 109017 Moscow, Russia}

\date{Accepted 000.
  Received 2007 June 000;
  in original form 2007 June 000}
\maketitle
\begin{abstract}
The rich oscillation spectra determined for the two
stars, $\nu$~Eridani and 12~Lacertae, present an interesting challenge to stellar modelling.
The stars are hybrid objects showing a number of modes at frequencies
typical for $\beta$ Cep stars but also one
mode at frequency typical for SPB stars.
We construct seismic models of these stars considering uncertainties
in opacity and element distribution. We also present estimate of
the interior rotation rate and address the matter of mode
excitation.

We use both the OP and OPAL opacity data and find significant
difference in the results. Uncertainty in these data remains a major obstacle in precise
modelling of the objects and, in particular, in estimating the
overshooting distance. We find evidence for significant rotation rate increase
between envelope and core in the two stars.

Instability of low-frequency g-modes was found in seismic models
of $\nu$~Eri built with the OP data, but at frequencies higher than those measured in the
star. No such instability was found in models of 12~Lac.
We do not have yet a satisfactory explanation for
low frequency modes. Some enhancement of opacity in the
driving zone is required but we argue that it cannot be achieved by
the iron accumulation, as it has been proposed.

\end{abstract}


{\section{Introduction}}

Transport of chemical elements is still not well understood aspect of stellar
interior physics. In the case of the upper main sequence there are
uncertainties regarding the extent of mixing of the nuclear
reaction products beyond the convective core, known as the
overshooting problem, as well as, regarding the survival of element
stratification in outer layers caused by  selective radiation
pressure and diffusion. Closely related is another difficult
problem of the angular momentum transport because there is a role
of rotation in element mixing. The upper main sequence pulsators,
$\beta$ Cephei stars, are potential source of
constraints on modelling the transport processes massive stars.
Recently,  Miglio et al. (2007b) and Montalb\'an et al. (2007)
studied sensitivity
of the oscillation frequencies in B stars to the rotationally induced mixing.
A comprehensive survey of properties of these variables was
published by Stankov \& Handler (2005).

Pulsation encountered in $\beta$ Cep stars frequently consists in
excitation of a number of modes, which differ in their probing properties. They
are found most often in evolved objects, where low order nonradial
modes have mixed character: acoustic-type in the envelope and
gravity-type in the deep radiative interior. Frequencies of such
modes are very sensitive to the extent of overshooting. Their
rotational splitting is a source of information about the deep
interior rotation rate. Pulsation is found both in very slow
($<10$~km~s$^{-1}$) and very rapid ($>200$~km~s$^{-1}$) rotating stars.
Thus, there is a prospect for disentangling the effect of rotation in element mixing.
First limits on convective overshooting and some measures of
differential rotation have been already derived from data on the
$\beta$ Cep stars HD 129929 (Aerts et al. 2003), $\nu$~Eri
(Pamyatnykh et al. 2004 (PHD), and Ausselooss et al. 2004).
Unfortunately, both objects are very slow rotators. To disentangle
the role of rotation in element mixing, we need corresponding
constraints for more rapidly rotating objects.

It seems that we understand quite well the driving effect
responsible for mode excitation in $\beta$ Cep stars. The effect
arises due to the metal (mainly iron) opacity bump at temperature
near 200 000 K. Yet, the seismic models of $\nu$~Eri, which
reproduce exactly the measured frequencies of the dominant modes,
predict instability in a much narrower frequency range than
observed. As a possible solution of this discrepancy, PHD proposed
accumulation of iron in the bump zone caused by radiation
pressure, following solution of the driving problem for sdB
pulsators (Charpinet at al. 1996). PHD have not conducted
calculations of the abundance evolution in the outer layers.
Instead, they adopted an {\it ad hoc} factor 4 iron enhancement
showing that it leads to a considerable increase of the
instability range but also allows to fit the frequency of the
troublesome p$_2$ dipole mode. However, recent calculations of the
chemical evolution made by Bourge et al. (2007) showed that the
iron accumulation in the $\beta$ Cep proceeds in a different
manner than in the sdB stars. Unlike in latter objects, in
photospheres of $\beta$ Cep stars the iron abundance is
significantly enhanced for as long as it is enhanced in the bump.
In fact, the photospheric enhancement is always greater (P.-O.
Bourge private communication).
 However, neither $\nu$~Eri nor
12~Lac shows abundance anomaly in atmosphere except perhaps for
the nitrogen excess, [N/O]=$0.25\pm0.3$, in the former object
(Morel et al. 2006). So there is no clear evidence for element
stratification and we are facing two problems: what is the cause
of element mixing in outer layers of this star, and how to explain
excitation of high frequency modes ($\nu>7$~cd$^{-1}$) and the
isolated very low frequency mode at $\nu=0.43$~cd$^{-1}$.

Unlike most of $\beta$ Cep stars, where frequencies of detected modes are
confined to narrow ranges around fundamental or first overtone of
radial pulsation, modes in $\nu$~Eri and 12~Lac are found in wide frequency ranges.
 Both stars are hybrid objects with simultaneously
excited low-order p- and g-modes (typical for $\beta$~Cep stars)
as well high-order g-modes (typical for SPB stars). The greatest
challenge is to explain how the latter modes are driven.
Our work focuses on these two stars, which
are best studied but not the only hybrid pulsators.
Other such objects are $\gamma$~Pegasi, $\iota$~Her, HD~13745, HD~19374
(De Cat et al. 2007). Very recently Pigulski \& Pojma\'nski (2008)
reported a likely discovery of five additional objects.

In the next section, we compare oscillation spectra and provide
other basic observational data for $\nu$~Eri and 12~Lac. Seismic
models of these two objects are presented in Section 3, after a
brief description  our new treatment of the overshooting. In the
same section, we give for both objects the estimate of rotation
rate gradient in the abundance varying zone outside core. Section
5 is devoted to the problem of mode excitation. Finally, in
Section 6, after summarizing the results, we discuss the matter of element mixing
in outer layers and what is needed for progress in B star seismology.

\vspace{4mm}
\section{Oscillation spectra and other observational data}

The multisite photometric (Handler et al. 2004, Jerzykiewicz et
al. 2005) and spectroscopic (Aerts et al. 2004) campaigns resulted
in detection of nine new modes in the  $\nu$~Eri oscillation
spectrum, in addition to four, which have been known for years.
Furthermore, data from these campaigns allowed to determine (or at
least constrain) the angular degrees for most of the modes
detected in this star. A schematic oscillation spectrum for this
star is shown in the upper panel of Fig.\,1. The angular degrees,
$\ell$, are adopted after De Ridder et al. (2004). In the lower
panel of Fig.\,1, we show the corresponding spectrum for 12~Lac.
All data, including the $\ell$-degrees, are from Handler et al.
(2006). The two spectra are strikingly similar. However, owing to
the appearance of two complete $\ell=1$ triplets, the one for
$\nu$~Eri is more revealing.

\begin{figure*}
\begin{center}
\includegraphics[width=170mm]{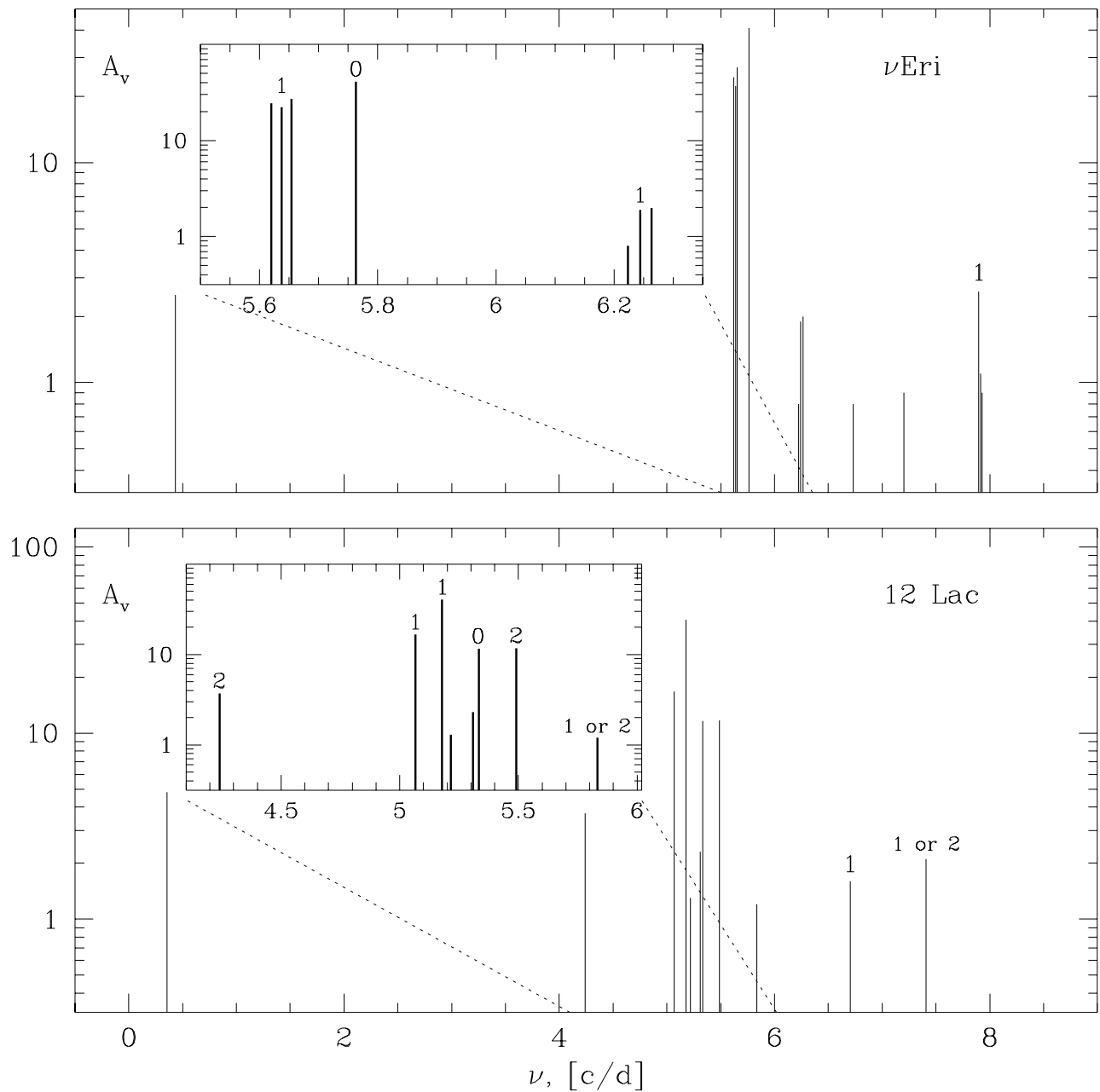}
\end{center}
\caption{Oscillation spectra of $\nu$~Eri and 12~Lac based on
Jerzykiewicz et al. (2005) and Handler et al. (2006) data,
respectively.  The numbers above the bars are the most likely
$\ell$-values, as inferred from data on amplitudes in four passbands of
Str\"omgren photometry.}
\end{figure*}

The mean parameters of the two objects are also similar. In Fig.\,2
we show their positions in theoretical H-R diagram with the errors
based on the parallax and photometry data (see caption for
details). In the same figure, we show also the evolutionary tracks
for selected seismic models of the stars which we will discuss in
the next section. Both objects have nearly standard Population I
atmospheric element abundances. In particular, Niemczura and
Daszy\'nska-Daszkiewicz (2005) quote the following values of the metal
abundance parameter [M/H]: $0.05\pm0.09$ for $\nu$~Eri and
$-0.20\pm0.10$ for 12~Lac. These numbers agree within the
errors with earlier determination by Gies \& Lambert (1992).
These authors also provide the $T_{\rm eff}$ values which are
left of the error box shown in Fig.\,2.
The adopted by us position in the H-R diagram
indicate that both stars are in advanced core hydrogen burning
phase.

The projected  equatorial velocity in both stars is low. Abt et
al.\,(2002) give 20 and 30 km~s$^{-1}$, for $\nu$~Eri and 12~Lac,
respectively, with the 9 km~s$^{-1}$ formal error. The
corresponding numbers quoted by Gies \& Lambert (1992) are
$(31\pm3)$~km~s$^{-1}$ and $(39\pm14)$~km~s$^{-1}$. For $\nu$~Eri,
the two values barely agree (within the errors) but are in
conflict with the seismic determination of PHD, who derive the
value of about  6 km~s$^{-1}$ for the equatorial velocity which we
believe to be a more reliable value. In Section 4, we will present
our refined seismic estimate of the rotation rate for this star
and the first such determination for 12~Lac.

\vspace{4mm}

\begin{figure}
\begin{center}
\includegraphics[width=85mm]{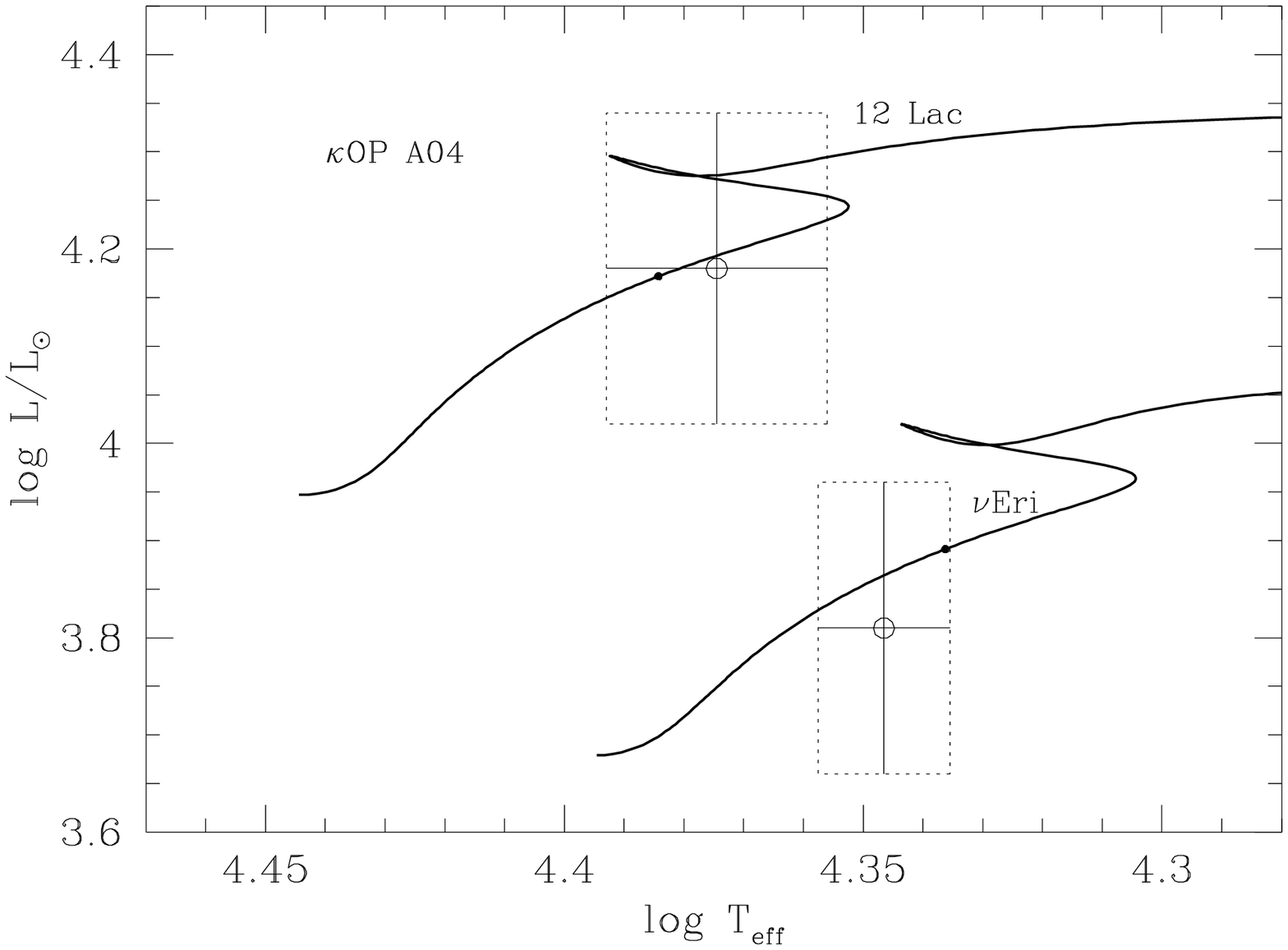}
\end{center}
\caption{ $\nu$~Eri and 12~Lac in the theoretical H-R diagram. The
the 1$\sigma$ error box for the first star is similar to that
adopted in PHD, except that it is moved downwards by 0.1 because
we do not include now the Lutz-Kelker correction to measured
parallax. For the second star, the error box is adopted after
Handler et al.\,(2006). The evolutionary tracks correspond to
seismic models (dots) calculated assuming no
overshooting and the initial hydrogen abundance $X=0.7$.
The $\nu$~Eri model has mass $M=9.625M_\odot$ and the heavy element mass fraction
$Z=0.0185$, and that of 12~Lac has $M=11.55M_\odot$ and
$Z=0.015$.}

\end{figure}

\vspace{4mm}
\section{Seismic models and interior rotation rate}

For specified  input physics, element mixing recipe, and heavy
element mixture, we need the four parameters: mass, the age, and
the initial abundance parameters, $(X,Z)$, to determine stellar
model, if dynamical effects of rotation are negligible. The latter
is very well justified for the two stars we are considering in
this paper. This might suggest that for testing the mixing
recipe, frequencies for more than four pulsation modes are
needed. This is not true because there are additional
(non-seismic) observational constraints on the models. On the
other hand, there are other uncertainties of modelling. The most
important concerns opacity.

In the approach adopted in this paper, like in PHD, we fixed $X$
at 0.7, as the changes within reasonable ranges do not
have any significant effect on frequencies. Furthermore, we
considered models in the expansion phase, so that instead of the
age we could use $T_{\rm eff}$ as the model parameter, which is
more convenient. There are two modifications in our modelling.
Firstly, in order to asses uncertainty in opacity, we constructed
models employing both OP (Seaton 2005) and OPAL (Iglesias \& Rogers 1996)
data and two heavy element mixtures, the traditional one of
Grevesse \& Noels (1993, hereafter GN93) and the new one of Asplund
et al. (2004, 2005, hereafter A04). Secondly, we allow for partial
mixing in the overshoot layer. In our reference model, we adopted
the OP opacities calculated for the A04 mixture and assumed no
overshooting.

To describe partial mixing, we use an additional adjustable
parameter. In the partially mixed layer, extending from the edge
of the convective core at the fractional mass $q_c$ to distance
$d_{\rm ov}=\alpha_{\rm ov}H_p(q_c)$, the fractional hydrogen
abundance was assumed in the form

\begin{equation}
X=X_c+(q-q_c)^w[a+b(q-q_c)].
\end{equation}

The input parameters are $w$ and $\alpha_{\rm ov}$. The  $a$ and
$b$ coefficients in the expression above were calculated from the
two input parameters by imposing continuity of $X$ and its
derivative at the top of the layer. The limit $w\rightarrow\infty$
corresponds to the standard treatment. We wanted to see, whether
this additional degree of freedom may help us to fit frequency of
a certain troublesome mode. In more general sense, it is important
to know whether asteroseismology may teach us about overshooting
more than only its distance.

In our seismic estimate of the interior rotation rate, we make a
similar assumption as PHD, that is in chemically homogeneous
envelope we assume a uniform rotation rate, $\Omega_e$, and allow
its sharp inward rise in the $\mu$-gradient zone. Here, we
additionally assume uniform rotation rate, $\Omega_c$, within the
convective core and a linear dependence of the rate $\Omega$ on
the fractional radius, $x$, in the adjacent $\mu$-gradient zone between $x=x_c$
and $x=x_{c,0}$, which is the fractional radius corresponding to
convective core mass at ZAMS. Thus, we use the following simple
expression for interior rotation rate

\begin{equation}
\Omega(x)=\left\{
\begin{array}{ll}
\Omega_c&\mbox{ if }x\le x_c\\
\Omega_c-(\Omega_c-\Omega_e){x-x_c\over x_{c,0}-x_c}&\mbox{ if
}x_c\le x\le x_{c,0}\\ \Omega_e&\mbox{ if }x\ge x_{c,0}
\end{array}
\right..
\end{equation}
Our aim is to determine $\Omega_c$ and $\Omega_e$ from measured or
inferred rotational splitting.

\subsection{$\nu$~Eridani}

Seismic models of this star constructed by PHD made use of the three
frequencies: 5.637, 5.763, and 6.224~cd$^{-1}$, which were associated
with ($\ell=1$, g$_1$), ($\ell=0$, p$_1$), and ($\ell=1$, p$_1$)
modes, respectively. Only the mean frequencies of the $\ell=1$
triplets are matter for seismic models. Individual frequencies probe
internal rotation. The frequency of 7.898~cd$^{-1}$ associated with
$\ell=1$ degree could not be reproduced with standard models. PHD
noted that the localized iron enhancement may remove the
discrepancy but, as we pointed out in the introduction, this is
unlikely. Ausseloos et al. (2004) succeeded  in fitting the four
frequencies only for rather unrealistic chemical composition
parameters and effective overshooting.

\begin{figure}
\begin{center}
\includegraphics[width=85mm]{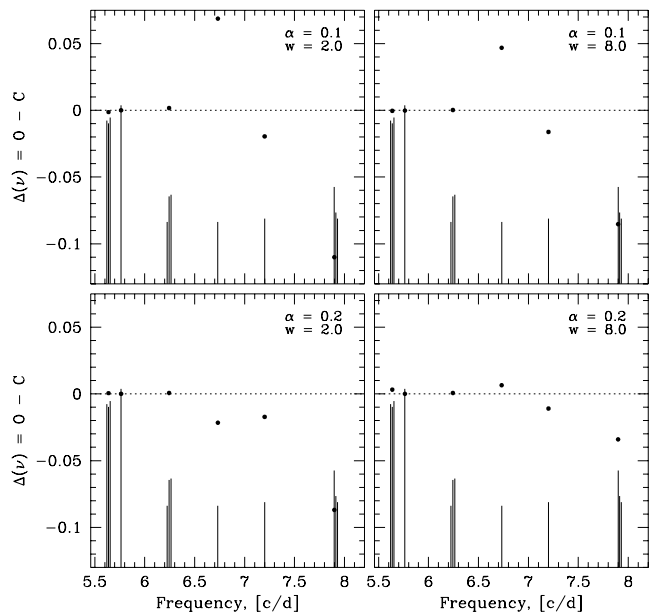}
\end{center}
\caption{Differences between observed and calculated
frequencies of $\nu$~Eri. All models have been calculated
with the OP opacity data and the A04 heavy element mixture. }
\end{figure}

In Fig.\,3, we show results of exploration of the two-parameter
freedom in description of the overshooting for fitting six
measured frequencies. All models are constructed to fit the three
lowest frequencies. Increase in $\alpha_{\rm ov}$ and/or in $w$
means more efficient overshooting and reduction of the O-C values
for the higher frequency mode. The price to pay for the improved
fit is shown in Fig.\,4. All models calculated with $\alpha_{\rm
ov}>0$ are outside the  $1\sigma$ error box in the H-R diagram.
Still, we regard efficient overshooting and cooler star as a
possible way of fitting the three high frequency modes. There is,
however, another problem regarding this part of the spectrum. The
two peaks at the highest frequencies cannot be interpret in terms
of remaining components of the $\ell=1, {\rm p}_2$  triplet
because the frequency separation should be similar to that in the
p$_1$ case and it is much smaller. We checked that there is no
other low degree mode in this vicinity, which could be
associated with any of the two peaks. Any one of them may be a
component of the triplet but certainly not both.

\begin{figure}
\begin{center}
\includegraphics[width=85mm]{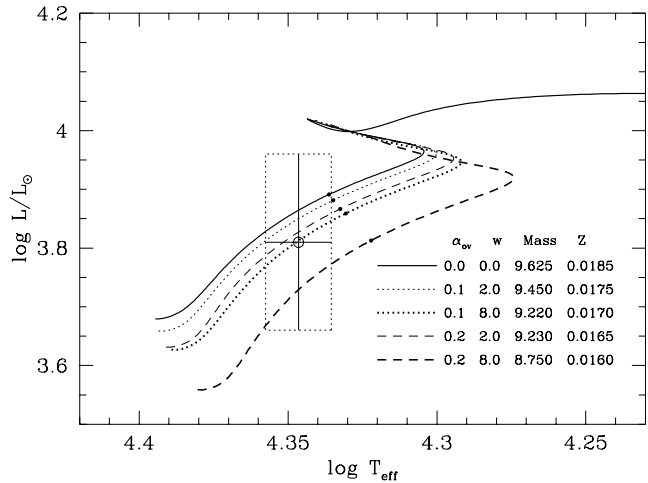}
\end{center}
\caption{ $\nu$~Eri in the theoretical H-R diagram. The the
1$\sigma$ error box is inferred from the observational data.
The evolutionary tracks correspond to seismic models (dots)
calculated for different values of the
overshooting parameters.}

\end{figure}

It is important to notice in Fig.\,3 the difference between modes
in the O-C changes. For the two highest
frequency modes, the shift in $w$ from 2 to 8 has a similar effect
to the shift in $\alpha_{\rm ov}$ from 0.1 to 0.2. For the mode at
6.732~cd$^{-1}$ the effect of the latter shift is much larger and
results in sign change of O-C. We identified this mode as $\ell=4,
{\rm g}_1$. This mode was detected only with the data from the
second photometric campaign (Jerzykiewicz et al. 2005) and no
$\ell$-value from the amplitude data has been assigned to it. In our
seismic models there is no lower degree mode near this frequency.
We may conclude that there is a prospect for getting additional
constraints on overshooting once we have enough frequency data.

As already noted by PHD, the peak at 7.2~cd$^{-1}$ may only correspond
to $\ell=2, {\rm p}_1$ mode. Its O-C values are obtained assuming
$m=0$ and they would be much closer to zero if $m=-1$ (retrograde)
identification is adopted. At this point, however, we are
reluctant to use this mode in seismic model construction. Higher
frequency modes are more sensitive to uncertainties in outer
layers and we know that some modification of the structure of this layer
is required because we have to solve the excitation problem.

In Fig.\,5, we show the effect of choice of opacity data and the
heavy element mixture on seismic models. We may see that the latter is
far less important. Models calculated with the OPAL data are
hotter and therefore may accommodate overshooting and stay within
the error box. However, relaying on the OPAL data will not help
the $\ell=1, {\rm p}_2$ mode frequency fit, as the entries in
Table 1 show. The smaller difference between measured and model
effective temperature, $\Delta\log T_{\rm eff}$, is always associated with greater O-C for
this mode. Therefore, it seems that if future measurements confirm the
current value of the effective temperature, the only way of
reconciling the seismic and non-seismic observable is an increase of
opacity in outer layers. In PHD this was accomplished by an
artificial enhancement of the iron abundance which had virtually
no effect on frequencies of the three modes used in the seismic
model but caused the desired -0.1~cd$^{-1}$ frequency shift of the $\ell=1, {\rm p}_2$
mode.

\begin{figure}
\begin{center}
\includegraphics[width=85mm]{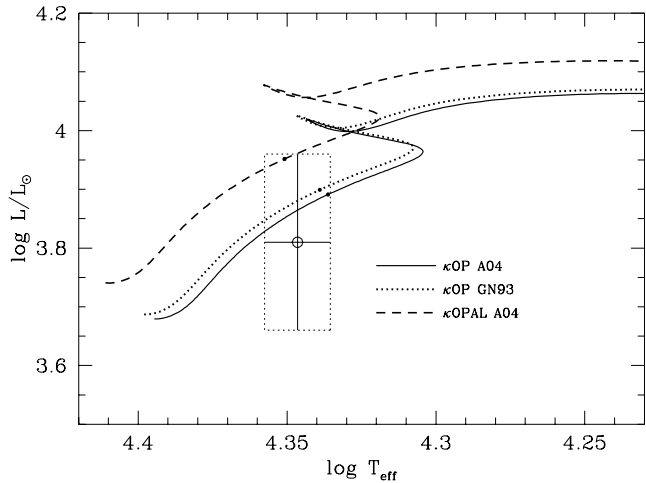}
\end{center}
\caption{ Influence of the choices of the opacity data and the
heavy element mixture on seismic models of $\nu$~Eri. The evolutionary
tracks were calculated assuming $\alpha_{\rm ov}=0$. }

\end{figure}

\begin{table}
\centering \caption{The O-C (in cd$^{-1}$) for the  $\ell$=1, $p_2$ mode
frequencies and the differences between observed effective
temperature and that of the seismic model, $\Delta\log T_{\rm
eff}=\log T_{\rm eff,obs}-\log T_{\rm eff,cal}$.
The observational uncertainty in effective temperature
is $\Delta\log T_{\rm eff}=0.011$.
Models with $\alpha_{\rm ov}>0$ were calculated with $w=8$.
$V_{\rm rot}$ is given in km~s$^{-1}$.}
\begin{tabular}{|ccccccccrrcccc|}
\hline
$\kappa$ & mixture & $\alpha_{\rm over}$ & O-C
& $\Delta\log T_{\rm eff}$ & $\Omega_c / \Omega_e$ & $V_{\rm rot}$\\
\hline
  OP   &  A04 & 0.0 & -0.127 & -0.0103 & 5.55 & 5.93 \\
\hline
  OP   & GN93 & 0.0 & -0.151 & -0.0075 & 5.36 & 5.95 \\
\hline
  OPAL &  A04 & 0.0 & -0.188 & -0.0044 & 5.36 & 5.99 \\
\hline
  OP   &  A04 & 0.1 & -0.085 & -0.0159 & 5.82 & 5.91 \\
\hline
  OP   &  A04 & 0.2 & -0.034 & -0.0244 & 5.78 & 5.93 \\
\hline
\end{tabular}
\end{table}

For a specified stellar model, the two parameters,
$\Omega_c$ and $\Omega_e$, describing internal rotation
may be directly determined from measured frequency splitting,
$S=0.5(\nu_+-\nu_-)$, of the two $\ell=1$ modes.
To this aim we use the relation
\begin{equation}
S=\int_0^1dx{{\cal K}(x)}{\Omega/2\pi},
\end{equation}
where ${\cal K}$ is the familiar rotational splitting kernel and
$\Omega$ is given by Eq.\,(2). The kernels for the these modes,
which were plotted by PHD, differ significantly and this is
essential for an accurate determination of $\Omega_c$ and
$\Omega_e$.

There are two differences relative to that work. Firstly, we now
have measured frequency of the $m=-1$ component of the higher
frequency triplet (${\rm p}_1$) and $\Omega_c$ denotes now the
rate in the convective core and not the mean value over the
$[0,x_{c,0}]$ range. In Table 1, we give the the values of the
$\Omega_c/\Omega_e$ ratio on the surface equatorial velocity,
$V_{\rm rot}=\Omega_eR$, calculated for the selected models. The
values differ only little. Certainly, the main uncertainty of our
inference is the $x$-dependence of its rate, which was assumed in
a very simplistic form (Eq.\,(2)).

\vspace{4mm} {\subsection{12~Lacertae}}

In Table 2 we list frequencies and angular degrees of the five
dominant modes in the oscillation spectrum of this star. The
frequencies and the $\ell$ values are from Handler et al.(2006)
analysis of their photometry data. The $m$ values were kindly
provided to us by Maarten Desmet and are based on his analysis of
line profile variations. He also provided us the value of
$V_e=52\pm5$~km~s$^{-1}$ for equatorial rotation velocity at the surface.

\begin{table}
\centering \caption{The frequencies and mode identifications
for 12~Lac.}
\begin{tabular}{|ccccccccrrcc|}
\hline
  Ident.   & Frequency   & $\ell$ & $m$    \\
           & (cd$^{-1}$) &        &        \\
\hline
  $f_1$    &  5.179034   & 1      & 1      \\
  $f_2$    &  5.066346   & 1      & 0      \\
  $f_3$    &  5.490167   & 2      & 2 or 1 \\
  $f_4$    &  5.334357   & 0      & 0      \\
  $f_5$    &  4.24062    & 2      & ?      \\
  $f_6$    &  7.40705    & 1 or 2 & ?      \\
  $f_7$    &  5.30912    & 2 or 1 & ?      \\
  $f_8$    &  5.2162     & 4 or 2 & ?      \\
  $f_9$    &  6.7023     & 1      & ?      \\
  $f_{10}$ &  5.8341     & 1 or 2 & ?      \\
\hline
\end{tabular}
\end{table}

The first attempt to construct seismic model of 12~Lac
(Dziembowski \& Jerzykiewicz 1999) concentrated on interpretation
of the very nearly equidistant triplet $(f_1,f_3,f_5)$. The
distances to the side peaks are -0.1558 and +0.1553~cd$^{-1}$.
Determination of the $\ell$ values by Handler et al. (2006) made
all proposed interpretations invalid. This is not a rotationally
split triplet. Equidistancy enforced by a nonlinear (cubic)
resonant mode coupling is also excluded by the $\ell$ values.
Thus, just a mere coincidence seems the only explanation.

Construction of seismic model of this star is more complicated
than that of $\nu$~Eri because it cannot be done separately of
estimate of the internal rotation rate. Therefore, in
this paper we limit ourselves to providing evidence for a
significant inward rise of the rotation rate. In this preliminary
analysis, we consider only models calculated with the A04 mixture
with $Z=0.015$ and without overshooting. Upon ignoring all effects
of rotation, we determined stellar mass and effective temperature by fitting
frequencies of the $\ell=0$, p$_1$(fundamental) and $\ell=1$,
g$_1$ modes to $f_4$ and $f_2$, respectively. We checked that
with the constraints on $L$ and $T_{\rm eff}$ there is no
alternative identification of radial orders of the two modes and
that models in the contraction phase are excluded.

The rotation rate of 12~Lac, as determined from spectroscopy and
implied by the $\ell=1$ mode splitting, is fast enough so that the
second order effects must be considered even at this preliminary
phase. Within the $\Omega^2$ accuracy we may write

\begin{equation}
\nu_m=\nu(0)+D_0(\Omega^2)+Sm+D_2(\Omega^2) m^2,
\end{equation}
where $\nu(0)$ is the frequency calculated ignoring all  effects
of rotation, $S$ is given in Eq.\,(3), the $D_0$  and $D_2$ terms include second
order effect of the Coriolis force and the lowest order effect of
the centrifugal force. In our treatment we rely on the formalism
described by Dziembowski \& Goode (1992) but simplified to the case of
the shellular ($\Omega=\Omega(r)$) rotation. In our treatment the
surface-averaged effect of the centrifugal force is incorporated in
evolutionary models and effects of the Coriolis force are included
in zeroth-order equation for oscillations. So the only
rotational frequency shift for nonradial modes
calculated by means of the perturbation theory is that caused
by centrifugal distortion. For radial modes we add only the shift
$4/3\nu_{\rm rot}^2/\nu_0$ caused by feedback effects of the toroidal
displacement induced by the Coriolis force.

We began assuming an uniform rotation and determined
its rate from the $f_2-f_1$ frequency separation using Eq.\,(4)
with coefficient calculated in the model reproducing the
the frequencies of $f_2$ and $f_4$ identified as corresponding to
$(n,\ell,m)=(-1,1,0)$ and $(1,0,0)$ modes. In this model the
frequency of the $(n,\ell,m)=(-1,2,1)$ mode is very close to
$f_3$. However, the inferred equatorial velocity of over 95 km~s$^{-1}$
is much larger than that determined from line profile changes.
An additional problem with this solution is the frequency
distance of nearly 0.3~cd$^{-1}$ between $f_5$ and the nearest
quadrupole mode, $(n,\ell,m)=(-2,2,2)$. We believe that these two
discrepancies exclude the hypothesis of uniform rotation.

\begin{figure}
\begin{center}
\includegraphics[width=85mm]{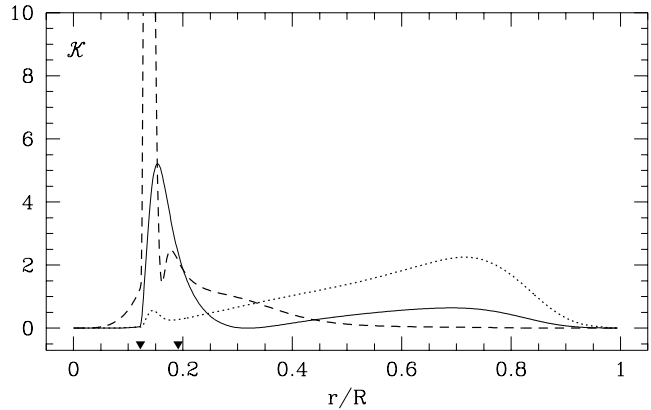}
\end{center}
\caption{ The splitting kernels for three noradial modes:
$\ell=1,\rm g_1$ (solid line), $\ell=2,\rm g_1$ (dotted line) and
$\ell=2,\rm g_2$ (dashed line).
Triangles mark boundaries of the $\mu$-gradient zone.}

\end{figure}

The splitting kernels for the three nonradial modes  plotted in
Fig.\,6 are very different. This means that there is a potential
for deriving significant constraints on differential rotation.
However, we have direct measurement of only one rotational
splitting, the others must be inferred through the frequency
fitting. At the equatorial velocity of rotation of about
100~km~s$^{-1}$, the mean effect of centrifugal force induces
frequency shifts exceeding 0.01~cd$^{-1}$. Thus, if we aim at such
a precision of the fit, we have to take into account differential
rotation in evolutionary models. This we leave for our future
work. In the present work, the effect of the differential rotation
described by Eq.\,(2) was rigorously treated only for calculation
of the rotational splitting while the mean shifts of the
multiplets induced the mean centrifugal force were determined
assuming uniform rotation corresponding to $\Omega_e$.

\begin{table}
\centering \caption{Models fitting the four dominant mode
frequencies ($f_1 - f_4$) at $\Omega_c=\Omega_e$ and
$\Omega_c=4.7\Omega_e$. The quantity $\Delta_5$ is the frequency
mismatch between $f_5$ and $\nu_{-2,2,2}$ in cd$^{-1}$.}
\begin{tabular}{|ccccccccrrcc|}
\hline
 $\Omega_c/\Omega_e$ & $V_{\rm rot}$ & $M/M_\odot$ &
 $\log T_{\rm eff}$ & $\log L/L_\odot$ & $\Delta_5$ & $m_{f_3}$ \\
\hline
  $1.0$ & 95.5  & 11.771 & 4.3878 & 4.194 & 0.113 & 1 \\
  $4.65$ & 47.1  & 11.777 & 4.3891 & 4.198 & -0.134 & 2 \\
\hline
\end{tabular}
\end{table}

We considered different values of the $\Omega_c/\Omega_e$ ratio
starting with 1 and used the $f_1-f_2$ frequency difference to
determine $\Omega_e$. Then, we adjusted the mass and effective
temperature to fit the frequencies of the four
dominant modes. As we may see in Table 2, there are two possible
$m$ values for the $f_3$ mode. With $m=1$ the fit is achieved at
$\Omega_c/\Omega_e=1$ and with $m=2$ at
$\Omega_c/\Omega_e=0.465.$ Data on the two seismic models are
summarized in Table~3.

We believe that our second model is more realistic because
its surface equatorial velocity agrees within $1\sigma$ with
observations. The mismatch between $f_5$ and calculated frequency
of the $(n,\ell,m)=(-2,2,2)$ mode is somewhat higher than in the
first model. However, it should be stressed that this mode is
most sensitive to detailed treatment of rotation in the inner
layers. We may see in Fig.\,6 that the splitting kernel, which is
similar to the mode energy distribution, for this mode is very
much concentrated in the $\mu$-gradient zone. Frequencies of such
modes are very sensitive also to overshooting.

\begin{figure}
\begin{center}
\includegraphics[width=85mm]{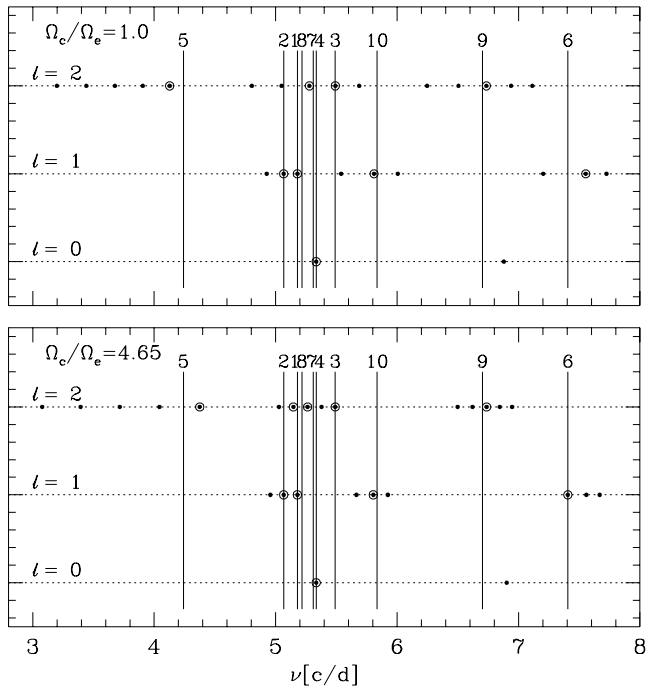}
\end{center}
\caption{Frequencies of high frequency modes detected in 12 Lac
(veritical lines) and frequencies off all $\ell\le2$ modes (dots) in the
two models (see Table 3). Identified modes are shown with
encircled dots. Note than for the model with uniform rotation
(upper panel), we have no identification for the $f_8$ peak. }

\end{figure}
Furthermore, as Fig.\,7 shows, the second model reproduces
closer frequencies of the remaining modes detected in 12 Lac. In
this model, $f_6$ may be identified as the p$_1$,~$\ell=1$,~$m=-1$
mode, while $f_7$ and $f_8$ may be, respectively, the $m=0$ and
$m=-1$ components of the g$_1$,~$\ell=2$ mode. It should be noted
that this identification for $f_7$ implies that this mode may be
coupled with radial mode associated with $f_4$, which is very
close. The slight frequency shift caused by this proximity (see
e.g. Daszy\'nska-Daszkiewicz et al. 2002) should be included in
the fine frequency fitting. There is a plausible identification of
$f_{10}$ as a p$_1$, $\ell=1$. The only troublesome peak is $f_9$
but only if indeed $\ell=1$, as Handler et al.(2006) proposed.
There would be no problem if it was $\ell=2$ because then the peak
could correspond to a fundamental ($n=0$) mode.

\vspace{4mm}
\section{Mode instability}

Satisfactory seismic models should account not only for measured
mode frequencies but also for their excitation. In practice, it
means that modes we want to associate with observed frequencies
should be unstable. As a  measure of mode instability
we use the normalized growth rate

 $$\eta={\frac{W}{\int_0^R
\left|\frac{dW}{dr}\right| dr}}~,$$
where $W$ is the global work integral. It follows from this
definition that $\eta$ varies in the range $(-1,+1)$ and it is $>0$
for unstable modes. The value of $\eta$ is more revealing than the usual growth
rate, because it is a direct measure of robustness of our conclusion regarding mode
stability.

In Fig.\,8, we show $\eta$ for $\ell=1$ and 2 modes calculated in
wide frequency ranges for our best models of the two stars. The
overall pattern of the $\eta(\nu)$ dependence in two stars is
similar. There is the instability range accounting for excitation
of most of the observed high frequency modes. There is also a
maximum of $\eta$ in the low frequency range not far from the
position of the low frequency modes detected in the two objects.
There is enough agreement to feel that we are on the right track
toward satisfactory seismic models. However, there is a need for
improvement.

Comparing our results for $\nu$~Eri with those presented by PHD,
we find a significant increase of the instability range.
Now, only the highest frequency modes near $\nu=7.9~{\rm cd}^{-1}$ and
the mode at $\nu=0.43~{\rm cd}^{-1}$ fall into stable ranges. This
change is a consequence of the usage of the OP data. That these
data lead to wider instability ranges than the OPAL data, has been already noted
before (see e.g. Miglio et al. 2007a).
With the OP data we have unstable $\ell=2$ high-order
g-modes ($n$=-14 to -20), but their frequencies are still too
high by factor at least 1.4. The measured frequency nearly
coincides with $\eta$ maximum for $\ell=1$ but the value is less than zero,
$\eta_{\rm max}=-0.14$. With the OPAL data we find maximum value
at nearly the same frequency but $\eta_{\rm max}=-0.33$. Perhaps,
there is still a room for improvements in opacity that would
render modes at the lowest frequency and the highest frequencies
unstable.

All modes detected in 12~Lac, except of that at $\nu=0.355~{\rm cd}^{-1}$,
occur in the unstable frequency range. However, the difficulty in
explaining excitation of the low frequency mode is even
higher than in the case of $\nu$~Eri. In part this is due to lower
mode frequency and in part to the lower heavy element mass fraction
adopted ($Z=0.015$).

\begin{figure}
\begin{center}
\includegraphics[width=85mm]{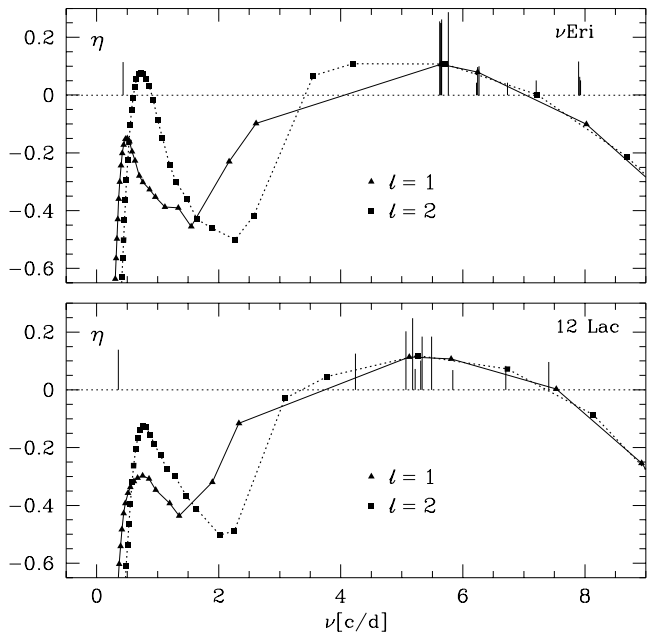}
\end{center}
\caption{ Normalized growth rate as a function of mode frequency in models of $\nu$~Eri (top) and
12~Lac (bottom) (the same as used in Fig.\,2). The frequencies of the detected modes are marked with vertical
lines starting at the $\eta=0$ axis. The length is proportional to the mode amplitude.}

\end{figure}

\vspace{4mm}
\section{Discussion}

We believe that our seismic models of $\nu$~Eridani yield a good
approximation to its internal structure and rotation but there is room for
improvement and a need for a full explanation of mode driving. We did not
fully succeed in the interpretation of the oscillation spectrum of $\nu$
Eridani with our standard evolutionary models and our linear nonadiabatic
treatment of stellar oscillations. The frequency misfit between
the high frequency peak to the $\ell=1, {\rm p}_2$ mode is much
reduced in models built with the OP opacity data allowing large
overshooting but such models are much cooler ($2\sigma$) than the mean colour of the star
implies. Such models also nearly avoid the driving problem. Similar
conclusions regarding models allowing large overshooting distance
were reached by Ausseloos et al. (2004). As for the consequences of
usage of the OP instead of OPAL opacity data for mode
instability, our finding agrees with more general observation of
Miglio et al. (2007a) that models using the former data predict
wider frequency ranges.

An assessment of the inward rise of the angular rate in the $\mu$-gradient zone
was made for both, $\nu$~Eri and 12~Lac, yielding the values
around five for the ratio of the core to envelope rate. The
surface equatorial velocities in the two stars are very
different. For the former object our seismic estimate gives about
6 km~s$^{-1}$. For 12~Lac, our estimate of about 50 km~s$^{-1}$ is less certain
because data on rotational splitting are much poorer but the value
agrees with estimate of Desmet (private communication), based on
his analysis of line profile changes. There is a large difference in rotation rate between
the two stars but unfortunately the accuracy of our modelling is insufficient
for addressing the question of the relation between the overshooting distance
and rotation.

The surface rotation rate in $\nu$~Eri is indeed very low and this
was the reason why PHD suggested that chemical element stratification in
this star may be sustained. Hovever, a closer look at the problem
reveals that the effect of rotation on element distribution should not
be ignored, even at the equatorial velocity of few km~s$^{-1}$.
As Bourge et al. (2007) showed for their $10 M_\odot$ model,
the two-fold excess of the iron abundance in the
driving zone is produced in the time scale comparable with main sequence life
time. The process is slower than in less massive objects
(Seaton 1999) but fast enough so that if it goes unimpeded a
significant excess of iron could be created.
The effect of rotation could be ignored if the time, $\tau_{\rm mc}$, for meridional circulation
to travel from the depth where most of iron is moved up
($T\approx4\times10^5$K)
to the bottom of the convective zone ($T\approx2\times10^5$K) around the iron opacity bump
is longer than the star age.
To estimate $\tau_{\rm mc}$ we use the well-known expression (see e.g. Tassoul 2000) for
the speed of the meridional flow

\begin{equation}
V_{\rm mc}={\epsilon R\over\tau_{\rm KH}}f(x)P_2(\cos\theta),
\end{equation}
where $\epsilon={\Omega^2R^3/GM}$,
$\tau_{\rm KH}={GM^2\over RL}$ is the Kelvin-Helmholtz time, and

$$f(x)={4\over3}\left({\partial\ln\rho\over\partial\ln T}\right)_p
{x^5\over1-\nabla/\nabla_{\rm ad}}\left(1-{\Omega^2\over2\pi G\rho}\right).$$

The last term, known as the Gratton-\"Opik term, is kept
though it is of higher order because it is important near the
surface where density $\rho$ is low, even if $\epsilon<<1$ and otherwise
linearization is valid. In our model of $\nu$~Eri, the the  Gratton-\"Opik term is -3.4
at the depth where most of iron is supposed to be pushed up.

\begin{equation}
\tau_{\rm mc}={\tau_{\rm KH}\over\epsilon}{\Delta x\over|f|}
\end{equation}

We evaluate $\tau_{\rm mc}$ at the layer where $T=4\times10^5$K.
and for $\Delta x$ use the distance to the bottom of the convective zone.
For the $\nu$~Eri model, we have $\epsilon=1.2\times10^{-4}$, $\tau_{\rm
KH}=5.9\times10^4$yr, $x_b=0.907$, $\Delta x=0.05$, $f=-15$. With this numbers, we get
from Eq.\,(6) $\tau_{\rm mc}=1.14\times10^6$y, which is much
less than the age of the star, $1.75\times10^7$y according to our seismic
model. Thus, even in this slowly rotating star the effect of rotation
cannot be ignored. For our best model of 12~Lac, we find $\tau_{\rm mc}$ by about four orders
less than in $\nu$~Eri. Yet, as we may see in Fig.\,8, the problem
with driving the low frequency mode is even greater than in the
case of $\nu$~Eri.

Meridional flow and/or turbulence developed through instability induced by
rotation are expected to prevent accumulation of iron in the
driving zone and in photosphere. This is a likely explanation of apparently normal chemical
atmospheric composition of $\beta$~Cephei stars, including such slow rotators as $\nu$~Eri.
This is also the reason to reject iron accumulation in the driving zone
as the solution of the driving problem, whose
solution must be searched in a different way.

We attach greatest hope for solution of the driving problem with improvements in
stellar opacity data. Thus, we would like to encourage further
effort in this field. Reliable opacity data are essential for
B star seismology. Data on mode frequencies in $\nu$~Eri and 12~Lac are abundant and accurate
enough for probing rotation and structure of the $\mu$-gradient
zone above the convective core. However, to answer open questions
regarding macroscopic transport of elements and angular momentum,
we need accurate microscopic input data, especially on opacity.
Only if we have measurements of  rotational splitting, our inference
on differential rotation does not rest on very precise model of
stellar structure, but such data are rare. Regarding observational
work, we view as most important new spectroscopic observations
of the two stars aimed at improving the value of the mean effective
temperature and leading to the $\ell$ and
$m$ identifications for a greater number of modes.

\section*{ACKNOWLEDGEMENTS}

This work was supported by the Polish MNiSW grant No~1~P03D~021~28.

\bsp

\end{document}